\documentclass[a4paper,twoside]{article}

\usepackage{epsfig}
\usepackage{subcaption}
\usepackage{calc}
\usepackage{amssymb}
\usepackage{amstext}
\usepackage{amsmath}
\usepackage{amsthm}
\usepackage{multicol}
\usepackage{pslatex}
\usepackage{apalike}
\usepackage{SCITEPRESS}

\begin{document}

\title{Wireless Sensor Network for {\em in situ} Soil Moisture Monitoring}

\author{
\authorname{Jianing Fang\sup{1}, Chuheng Hu\sup{2}, Nour Smaoui\sup{3}, Doug Carlson\sup{4*}, Jayant Gupchup\sup{5*}, Razvan Musaloiu-E.\sup{*}, Chieh-Jan Mike Liang\sup{6*}, Marcus Chang\sup{7*}, Omprakash Gnawali\sup{2}, Tamas Budavari\sup{8}, Andreas Terzis\sup{*}, Katalin Szlavecz\sup{1}, and Alexander S. Szalay\sup{9}}

\affiliation{\sup{1}Department of Earth \& Planetary Sciences, Johns Hopkins University, 3400 N Charles Street, Baltimore, USA}
\affiliation{\sup{2}Department of Computer Science, Johns Hopkins University, 3400 N Charles Street, Baltimore, USA}
\affiliation{\sup{3}Department of Computer Science, University of Houston, 4800 Calhoun Rd, Houston, Houston, USA}
\affiliation{\sup{4}CEVA, Inc., 15245 Shady Grove Rd, Rockville, USA}
\affiliation{\sup{5}Microsoft Corporation, One Microsoft Way, Redmond, USA}
\affiliation{\sup{6}Microsoft Research, No. 5 Danling Street, Beijing, China}
\affiliation{\sup{7}Arm Ltd., 5707 Southwest Pkwy 100, Austin, USA}
\affiliation{\sup{8}Department of Applied Mathematics \& Statistics, Johns Hopkins University, 3400 N Charles Street, Baltimore, USA}
\affiliation{\sup{9}Dept. of Computer Science and Physics \& Astronomy, Johns Hopkins University, 3400 N Charles Street, Baltimore, USA}

\email{jfang25@jhu.edu, chu29@jhu.edu, nsmaoui@central.uh.edu, carlsondc@gmail.com, 
    Jayant.Gupchup@microsoft.com, razvanm@cs.jhu.edu, liang.mike@microsoft.com, marcus.chang@arm.com, gnawali@gmail.com, budavari@jhu.edu, aterzis@gmail.com, szlavecz@jhu.edu, szalay@jhu.edu
    }
    
\affiliation{\sup{*}Previous members of the team project. Work performed while at Johns Hopkins University.}

}


\abstract{We discuss the history and lessons learned from a series of deployments of environmental sensors measuring soil parameters and CO$_2$ fluxes over the last fifteen years, in an outdoor environment. We present the hardware and software architecture of our current Gen-3 system, and then discuss how we are simplifying the user facing part of the software, to make it easier and friendlier for the environmental scientist to be in full control of the system. Finally, we describe the current effort to build a large-scale Gen-4 sensing platform consisting of hundreds of nodes to track the environmental parameters for urban green spaces in Baltimore, Maryland.}

\onecolumn \maketitle \normalsize \setcounter{footnote}{0}\vfill

\section{Introduction}
\label{sec:introduction}
Soil is a semi-aquatic habitat harboring a huge diversity of terrestrial as well as aquatic organisms. The amount and availability of soil water affect survival and activity of organisms from bacteria to macrofauna to plants directly, but also indirectly, by redistributing nutrients or toxic substances. Consequently soil water content is a main driver of belowground biological activity. Pathways and rates of complex biogeochemical processes can shift dramatically especially if the soil dries out or becomes waterlogged. For instance the same ecosystem can switch between being a methane sink or source depending on fluctuations of water table or precipitation patterns.  

Soil is inherently spatially heterogeneous in all three dimensions and at many spatial scales. Young and Crawford (2004) called soils “the most complicated biomaterials on the planet” \cite{Young04}. On field scale high degree of patchiness exists even in seemingly uniform landscapes \cite{Robertson94}. Soil physical, chemical and biological characteristics vary even more in fragmented landscapes. An extreme example of this is the urban-suburban environment with highly modified surface topography and soil conditions. Land use-land management decisions (construction, plant cover, irrigation regimes) are often made on a parcel level to the extent that it overrides the characteristics of the underlying natural soil \cite{Pouyat10,Pickett10}. To obtain spatial soil moisture data with modeling approaches, such as using digital elevation models to produce topographic moisture index, proved to be challenging in such disturbed environments \cite{Tenenbaum06}. 

For soil ecology research, understanding soil moisture conditions on a field scale is paramount to identify biogeochemical activity ‘hot spots’ and ‘hot moments’, and to explain spatial distribution of soil organisms. Spatially explicit approach to study soil communities is relatively new, but increasingly recognized as a significant tool in explaining high soil biodiversity \cite{Ettema02,Chust03}.  One approach to collect in situ, spatially explicit, continuous soil data is to use sensor systems, more specifically wireless sensor networks (WSN hereafter). This technology emerged in the late 1990’s \cite{Cerpa01}, and promised inexpensive, hands-free, low-cost, and low-impact environmental data collection — an attractive alternative to manual data logging — in addition to providing considerably more data at finer spatial and temporal granularities. Since originally complete off-the shelf solutions were not available, and both the sensors themselves and the related data systems needed customization, we, a group of soil ecologists, physicists and computer scientists started building our own WSN  to monitor the soil ecosystem. 

Our project, dubbed LifeUnderYourFeet (LUYF) developed three generations of hardware and software, and deployed the systems in various habitats including tropical and temperate forests, agricultural fields and a desert. In this paper we discuss our efforts and experiences in designing and deploying an end-to-end system using wireless sensor networks to monitor soil conditions. 

\section{History of Our WSN}

\begin{figure}[ht]
\centering
\includegraphics[width=0.45\linewidth]{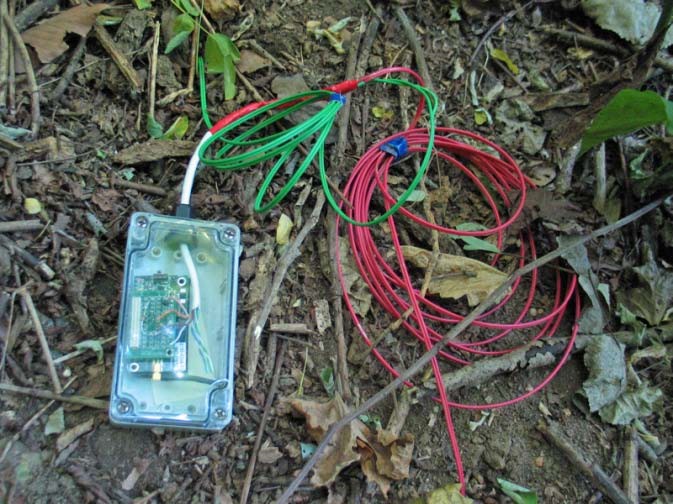}
\includegraphics[width=0.45\linewidth]{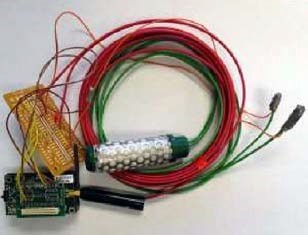}
\caption{\small Our first platform, based on the MicaZ in Hammond boxes.}
\label{fig:gen1}
\end{figure}

{\bf Generation 1.} Since our project started in 2005 we have gone through three generations of data collection equipment. The first wireless devices were manufactured by CrossBow Inc. These MicaZ ``mote'' (Figure \ref{fig:gen1}) contained processing, storage, and communication capabilities in a small form factor (matchstick box size). Besides two simple on-board sensors for temperature and light, five additional external analog sensors could be connected simultaneously to the mote via a separate fan-out board. In practice, this board turned out to be very fragile and was one of the reasons we abandoned this platform. The device’s on board flash was capable of  storing thousands of sensor measurements for later retrieval. However, the MicaZ motes were quite expensive costing close to \$200 each at the time.

\begin{figure}[ht]
\centering
\includegraphics[width=0.75\linewidth]{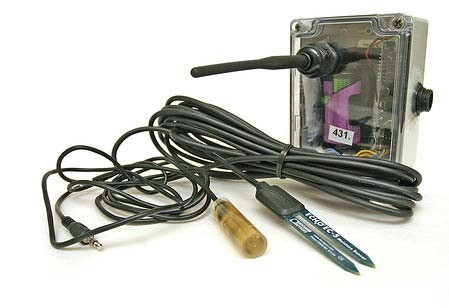}

\caption{\small Our Gen-2 platform, based on the Telos mote.}
\label{fig:gen2}
\end{figure}

For the soil measurements, we first built our own temperature sensors by placing small thermistors into plastic vials, about 1.75 in long, and filling them with epoxy. While the construction was moderately labor intensive, the material cost was less than \$5 (Figure \ref{fig:gen2}). For soil moisture we first experimented with the Watermark moisture sensor by Irrometer. The performance of these sensors rapidly degraded with time, thus we switched to the ECH$_2$O EC-5 moisture sensors by Decagon. We used two AA batteries to power both the MicaZ mote and its sensors.

In order to protect the motes from the environment we build our own enclosures out of watertight IP67-rated Hammond enclosures. While this enclosure setup survived the “bathtub test”, i.e. remained dry for extended periods under water in a bathtub, the prolonged exposure to the changing weather conditions along with the holes that we drilled for the sensor cables, even though ample sealant was used, was enough to compromise its water-tight feature.

\vspace{4pt}
\noindent {\bf Generation 2.} In 2007 we switched to a new wireless device, the Tmote Sky from MoteIV, but kept the custom-built temperature sensors and the Decagon moisture sensors (Figure \ref{fig:gen2}). The Tmote Sky is in many ways quite similar to the MicaZ, but considerably less expensive. This platform offered four times the measurement fidelity and twice as much space for storing samples. We designed and built an expansion board that allowed us to handle four external analog sensors. 

Besides the external sensors, this mote also had a built in digital humidity and temperature sensor by Sensirion (as opposed to the analog temperature sensor on the MicaZ) and two light sensors, one for visible light and the other one for photosynthetically active radiation. However, as with the MicaZ, the light sensors were quite sensitive, so under normal daylight they were either entirely ``off'' or ``on''. 

For the enclosure we still used a Hammond box, but in order to increase the range of the radios, we attached an external antenna (5 dBi gain), directly connected to the mote, but extending outside the box. The hole for the antenna was sealed with silicone based caulk. The expansion board was put in its own case and filled with insulation foam after the moisture and temperature sensors had been attached. We also upgraded the power supply to a single D-sized Li-SoCl2 battery that increased the capacity by an order of magnitude. 

The new platform was much more stable and professional, but not without its weak points. First, the sensor connection was quite expensive. The connectors and the cable for the expansion board cost almost \$30. Furthermore, assembling the whole setup was cumbersome, requiring considerable manual labor. While there were almost no cases in which we had standing water inside the enclosures, moisture tended to accumulate over time, and we placed moisture absorbing silicone beads inside each enclosure to mitigate this problem. 
\begin{figure}[ht]
\centering
\includegraphics[width=0.45\linewidth]{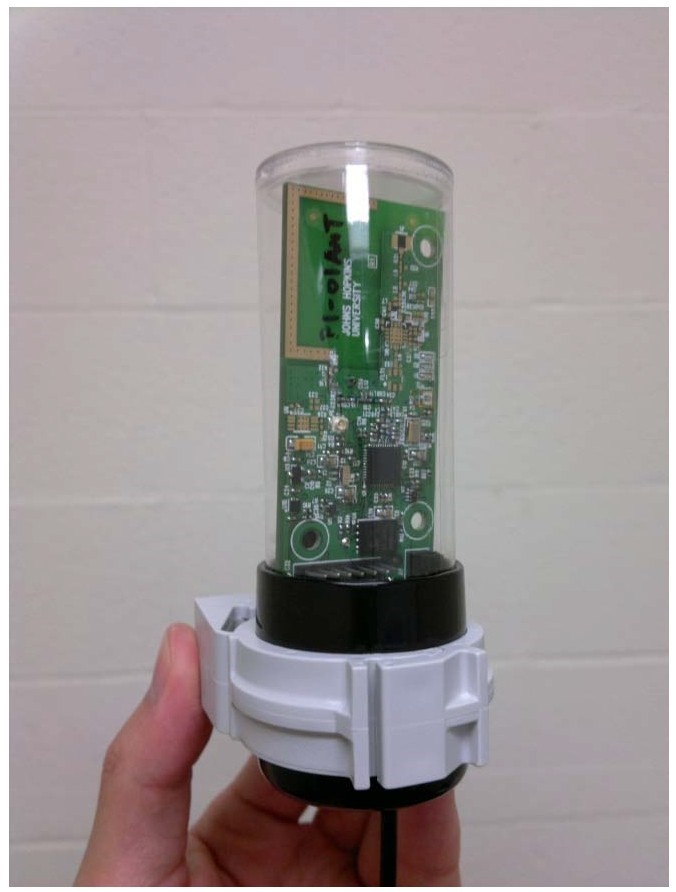}
\caption{\small Our Gen-3 platform, based on our PCB design.}
\label{fig:gen3}
\end{figure}

\vspace{4pt}\noindent{\bf Generation 3.} In 2013 we have designed and started deploying our third generation sensing platform, the Breakfast Suite, consisting of the Bacon mote (Figure \ref{fig:gen3}),
and the Toast multiplexer module. These were all designed in house and manufactured by an OEM in Hungary. This new mote encapsulates our experiences over the past seven years and is specifically designed for soil monitoring at different scales. With the same form factor as the Tmote Sky, the Bacon mote has five times the processing power and eight times the storage space. Moreover, its average power consumption is an order of magnitude lower than the Tmote sky and the manufacturing cost for thousands of units was only \$20, in 2012. To achieve large scalability, both in terms of coverage area and sensor density, we increased the communication range of the Bacon mote with more efficient radios and designed a new expansion board, the Toast sensor board, capable of reading eight external sensors at a time. Multiple Toast boards can be attached to a single Bacon mote at a time thereby facilitating different sensor densities and more flexible sensor compositions. Furthermore, we have added a much simpler multiplexer board (miniToast) to the palette, which is code compatible with the Toast, but contains an on-board high precision thermistor and a port for additional analog sensor.

\section{History of Deployments}
\label{sec:history}
Roughly parallel to the evolution of hardware platforms over the years, our deployment architecture went through several changes as we identified weak points and expanded our usage scenarios. At a high level, we first increased reliability by adding a permanently-powered ``basestation'' collection mote (with Internet access), then worked to maintain reliability while decoupling network performance from the presence of the basestation.

\begin{figure}[ht]
\centering
\includegraphics[width=0.85\linewidth]{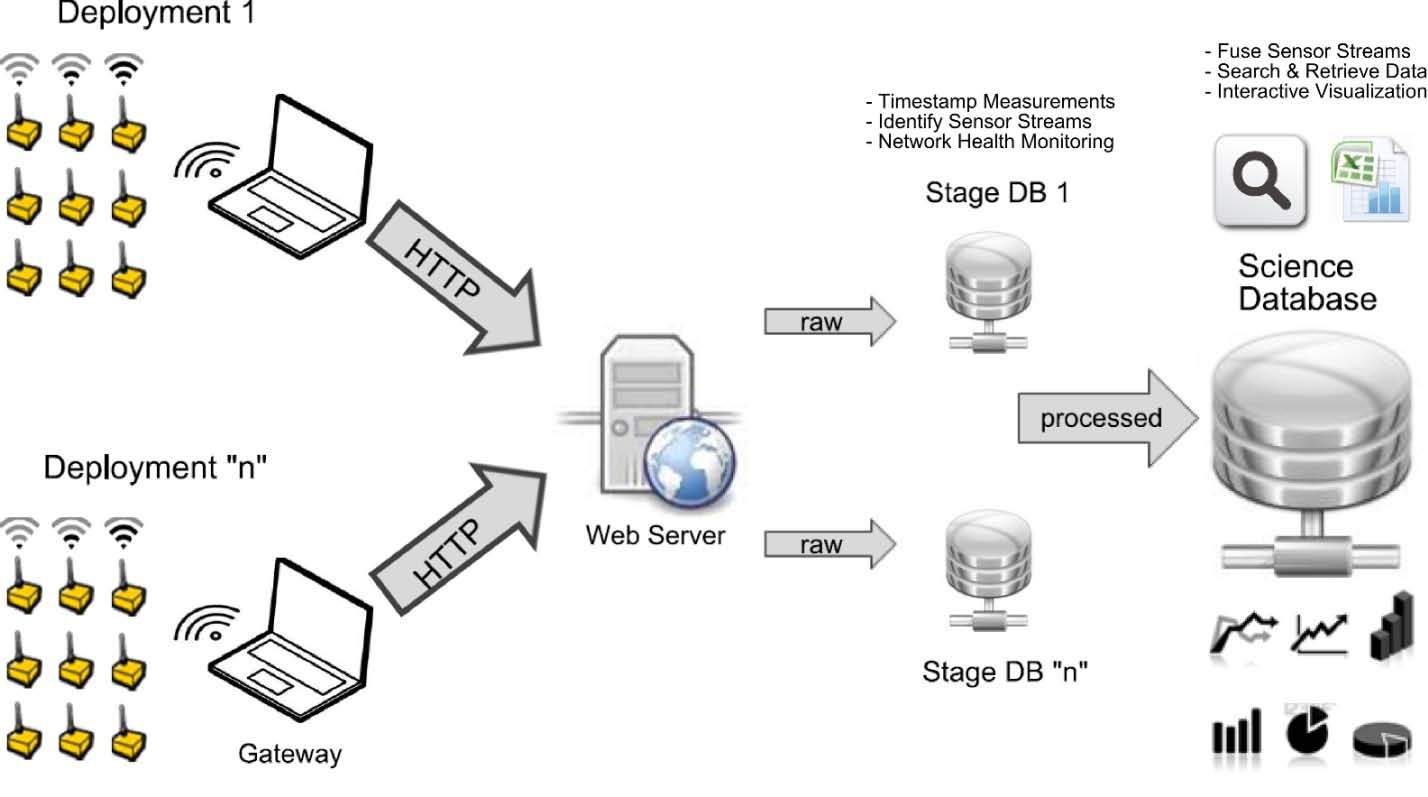}

\caption{\small System architecture of Generation 2 wireless sensor network.  Many deployments of sensing nodes send data through a Gateway computer to a web server. The raw analog sensor measurements are assigned timestamps converted to physical units, and screened for faults in the Stage-database.  This data is mapped to geographical locations and the data is organized for retrieval in the science database.}
\label{fig:architecture}
\end{figure}

\noindent{\bf Early Deployments.} The early deployments (Figure \ref{fig:architecture}) had no persistent infrastructure: motes logged data internally and a field visit was required to collect their data over a wireless connection to a laptop. While this approach was more convenient than using wired data loggers, it was still a tedious manual process. More importantly, due to the lack of a reliable clock on the motes, periodic references to some external time source were required to map measurements to points in time. If a mote became unstable (due to low battery, condensation, or software faults), it could record data that lacked time references. This problem was addressed by using the on-board light sensors to estimate the date (by matching the estimated length-of-day to a physical model of the Earth’s movement around the Sun) and local noon (the midpoint between dawn and sunset). Even though the precision of this approach was limited to a few minutes, it was still adequate for the relatively low sensor sampling rate we used (one sample every 5-10 minutes) \cite{Gupchup09}.

\begin{figure}
\centering
\includegraphics[width=1\linewidth]{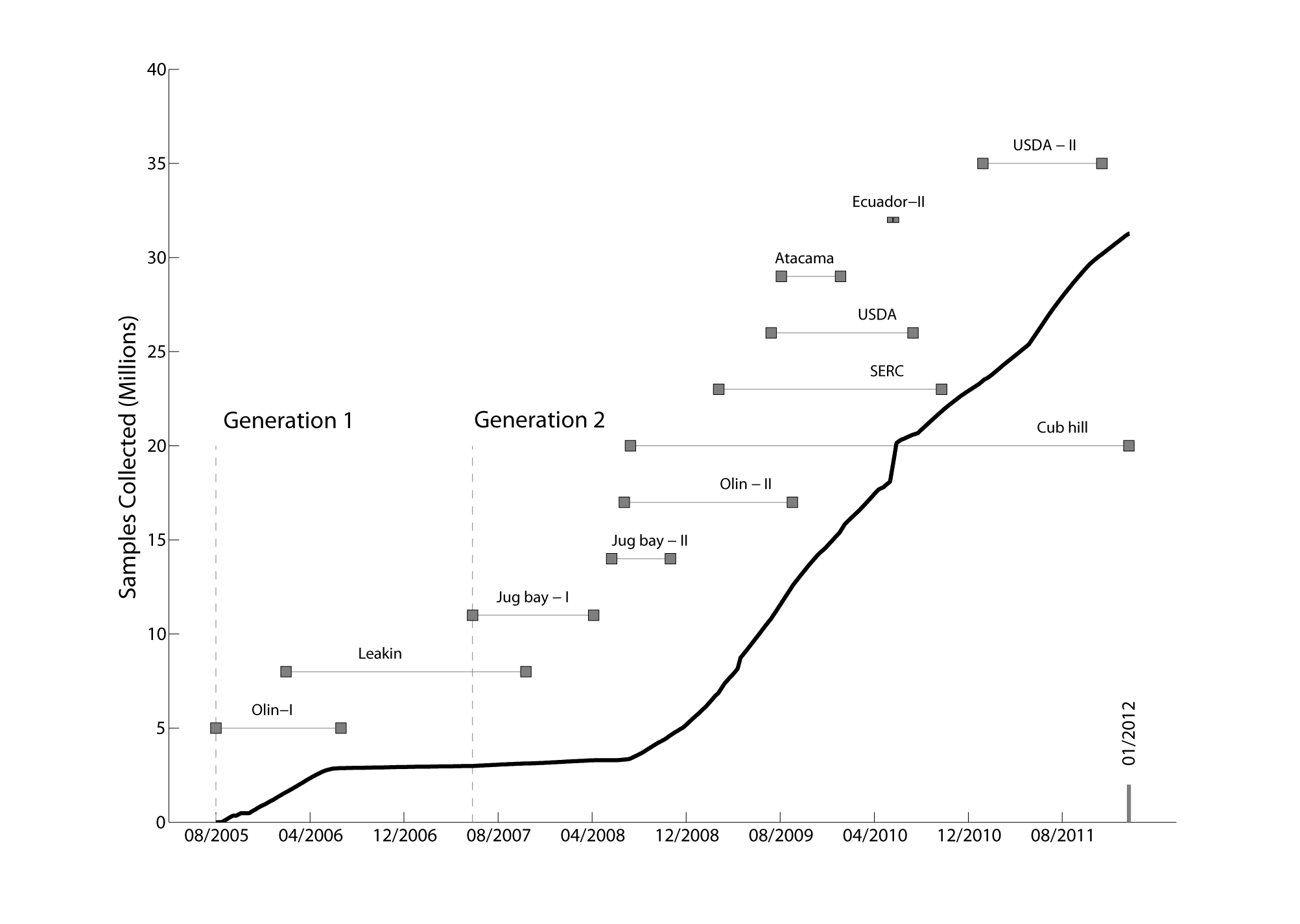}
\caption{\small Cumulative number of soil moisture samples collected in the LifeUnderYourFeet project. Major hardware changes are marked with vertical lines, while horizontal lines show the start and end of deployments.}
\label{fig:samples}
\end{figure}

The first deployment occurred on the Johns Hopkins University campus in a wooded area and lasted for seven months.  Subsequent deployment took place in urban forests in Baltimore (Leakin park, Cub Hill), deciduous forests in the Mid-Atlantic Coastal Plains (Smithsonian Environmental Research Center, SERC, Jug Bay), agricultural fields (USDA), tropical rainforests (Ecuador), and high altitude deserts (Atacama) (Figure \ref{fig:samples}). The objectives of collecting soil moisture and temperature data in these projects were diverse and included measuring soil carbon cycling, monitoring turtle overwintering behavior and egg incubation, predicting soil denitrification activity and testing the system at low barometric pressure and high UV exposure. 

\begin{figure}[ht]
\centering
\includegraphics[width=0.55\linewidth]{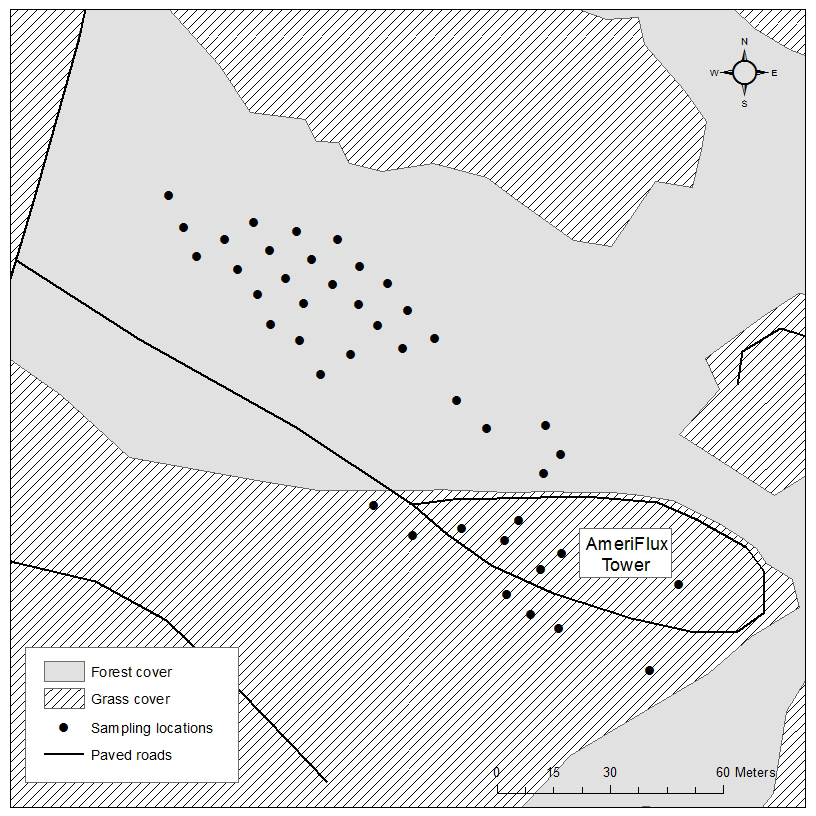}
\caption{\small Land cover map of the Cub Hill site. Each sampling location had one mote with two soil temperature and two soil moisture sensors attached. Measurements were taken at 10 and 20 cm depths. }
\label{fig:cubhill}
\end{figure}

\vspace{4pt}\noindent{\bf The Cub Hill Project.} The largest and longest deployment was in an urban residential area, Called Cub Hill. The Cub Hill site (39.412507 N,-76.520903 W) is located in Parkville, Baltimore County, MD, 14 km north from the center of Baltimore City.  The site has several ongoing studies related to the Baltimore Ecosystem Study LTER (www.beslter.org) including long-term anthropogenic effects focused on the atmosphere, soils, hydrology, and vegetation \cite{Pickett08,Pickett10}.  A permanent urban CO$_{2}$ flux tower operates at the juxtaposition of forest and residential areas (Figure \ref{fig:cubhill}). To the North and West of the tower is a poplar-oak-hickory stand with a canopy height of 20-26 meters.  To the South is a mix of medium density residential areas made up of several subdivisions built in the 70’s and 80’s.  Detailed soil and land use mapping has been carried out in the 1 km$^2$ footprint of the area \cite{Ellis06,Yesilonis18}. 

The majority of motes were deployed in the two main land cover types, forest and grass, but a small number of sensors were installed in planting beds, covered with mulch or English Ivy. At each sampling location soil temperature and soil moisture were measured at 10 and 20 cm depths, thus at maximum capacity a total of 106 soil moisture values were collected at each sampling. Measurements were taken every 10 minutes.   

Maintenance ended in June 2011 and the project terminated in 2012, at which point we brought all sensors back to the lab. We were interested in the performance of the soil moisture sensors after being in the field for several years. First, all sensors were inspected for physical damage and categorized as follows: no visible damage, minor cracks in probe, and major damage due to large cracks with internal wires exposed. Laboratory tests were performed in a 10 gal Sterilite storage bin using mineral soil from one of our field sites that had been previously dried and sieved for homogeneity. The soil was gradually remoistened and mixed to approximately 0.2 volumetric water content (VWC). For testing, batches of eight sensors were fully inserted into the soil and their moisture readings were recorded every 1.5 sec (approx.) for two minutes using custom developed data loggers (www.lifeunderyourfeet.org). Sensors that provided measurements that were noticeably different than what others were reading often were tested several times. Sensors were assessed based on how similar their mean VWC was to the grand mean of all sensor VWC means (excluding obviously bad values) and how tight their sampling distribution was according to their coefficient of variation (CV).  Sensors were categorized as “good” if their mean VWC was within 0.03 VWC of the grand mean of all sensor means (based on the 3\% error rate specified by Decagon Devices for mineral soil) and their CV was less than 0.03.   “Fair” sensors had a mean VWC within 0.03 VWC of the grand mean, but a CV greater than 0.03. “Bad” sensors fell out of the allowable range for both criteria.

The Cub Hill experiment in Baltimore City studied the urban/forest gradient, near an AmeriFlux tower. It exemplifies our second generation of deployments. These added a persistent basestation mote (attached to an Internet-connected PC) that could automatically download data periodically and transfer it to a database (Figure \ref{fig:architecture}). These deployments also used the Koala \cite{Musaloiu08} low power collection protocol to enable multi-hop downloads (where the network cooperates to relay data for motes which are out of the communication range of the basestation). While regular contacts with the basestation helped to obviate the need for a reliable on-board clock, there were still a few periods during which power outages at the deployment site, heavy snow, or general poor network conditions led to timestamp losses. 

In order to support deployments in remote regions (lacking permanent power), we made a few modifications to the Gen-2 software. We implemented a system where motes exchange local time references with each other which, when collected with the data, can be used to map measurements to their correct point in time, even when there is no basestation for weeks \cite{Gupchup10}. We added several GPS-equipped motes to the network which served as reliable time references. Finally, we implemented data compression on the motes, so that even if conditions prevented them from being accessed at all once deployed, they would have enough space to buffer all their measurements for the entire deployment duration \cite{Carlson10}. 

\vspace{4pt}\noindent{\bf Smithsonian Environmental Research Center.} The Gen-3 software decoupled the energy requirements of individual motes from the overall network size by segmenting deployments into semi-autonomous patches. This multi-tiered network approach is a critical step towards enabling large-scale deployments. Our largest Gen-3 deployment was at the Smithsonian Environmental Research Center in Edgewater, Maryland. The system monitored air and soil temperature and soil moisture in deciduous forest stands. For the Mid-Atlantic Region of North America, climate models predict increased amount of rainfall, distributed more unevenly throughout the year. Under constructed rainout shelters we manipulated the frequency and intensity of rainfall, and followed the fate of carbon using isotopically enriched leaf litter.  The pilot experiment lasted for five months. There were two patches (old and young forest), and two manipulations (wet and dry) in each stand. Each rainout shelter was equipped with three sensor assemblies, collecting data from 24 sensors. Another two were deployed randomly in the forest to collect background data. Over the course of five months we collected over 90,000 data points.

\section{Lessons Learned}

The LifeUnderYourFeet project has collected over 400MB of raw sensor data, approximately 20\% which were soil moisture data (Figure \ref{fig:samples}).  The Cub Hill deployment itself collected 100,745,909 samples, out of which 39,471,983 and 17,627,640 were environmental and specifically soil moisture data, respectively. Below we present a selected set of these results, both on the performance of the WSN and the sensors and on soil conditions. Other aspects of the data collection and analysis are reported in \cite{Musaloiu08,Gupchup10,Savva12}.

\vspace{4pt}\noindent
{\bf Performance of the wireless sensor network}
\vspace{4pt}

\begin{figure}
\centering
\includegraphics[width=0.9\linewidth]{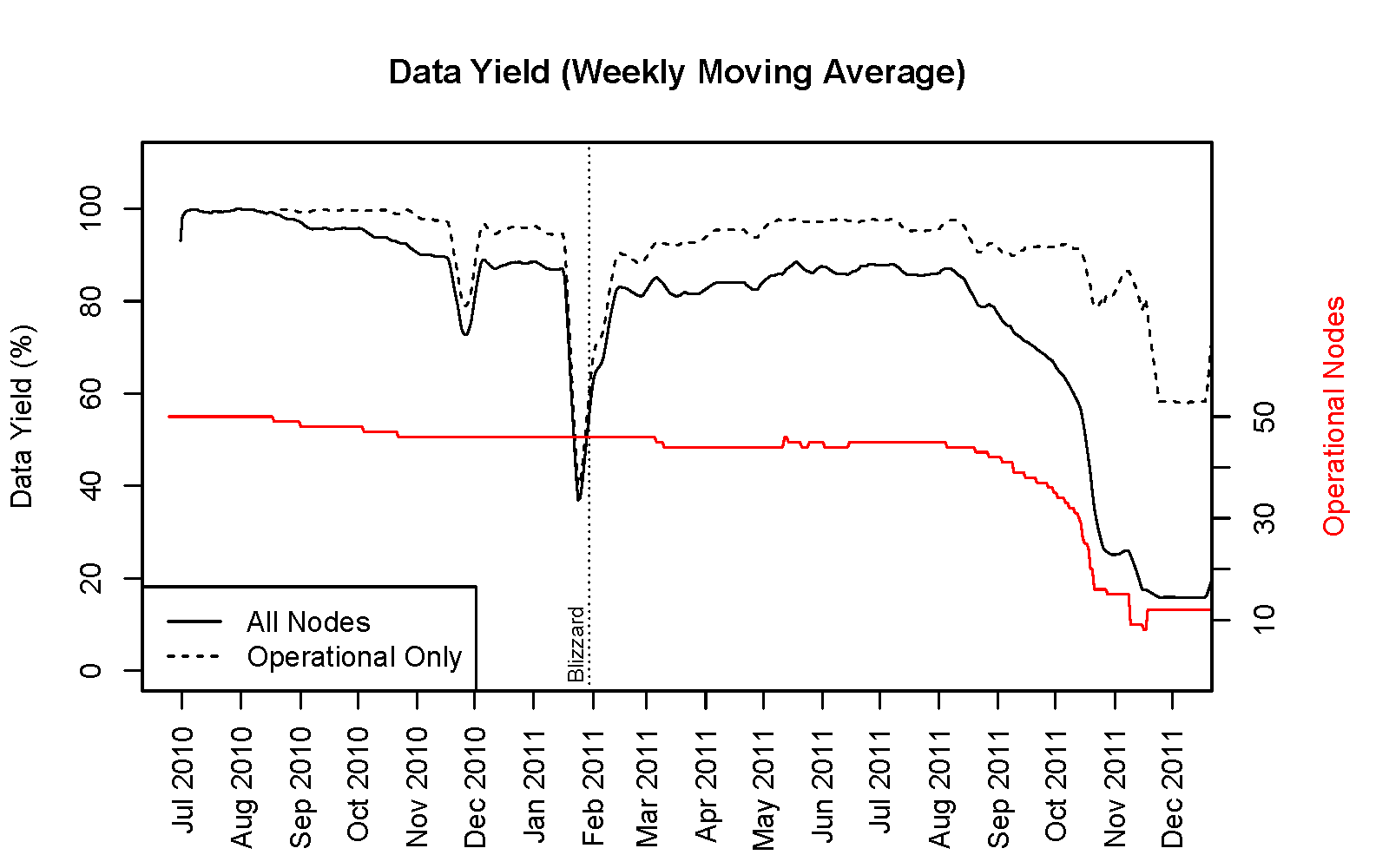}
\caption{\small Daily data yield (as a fraction of expected data yield) for the Cub Hill deployment. Starting point (July 2010) indicates a system-wide battery replacement. The solid line indicates how much data was recovered as a fraction of the entire 50-node network's expected yield, while the dashed line only considers the remaining operational nodes.    }
\label{fig:fig7}
\end{figure}

\noindent
Figure \ref{fig:fig7} shows how the number of operational nodes and total data yield (timestamped and fault-free data) varied over an 18-month maintenance-free period (Figure \ref{fig:fig7}). All nodes were installed with fresh batteries at the end of June 2010 and were not serviced again. While the yield declined slowly as devices failed, the remaining nodes continued to work well even as the network degraded around them. It's not until November of 2011 that so few nodes remain in operation that it became difficult to reach them and obtain time references for them.

\begin{figure}[ht]
\centering
\includegraphics[width=0.8\linewidth]{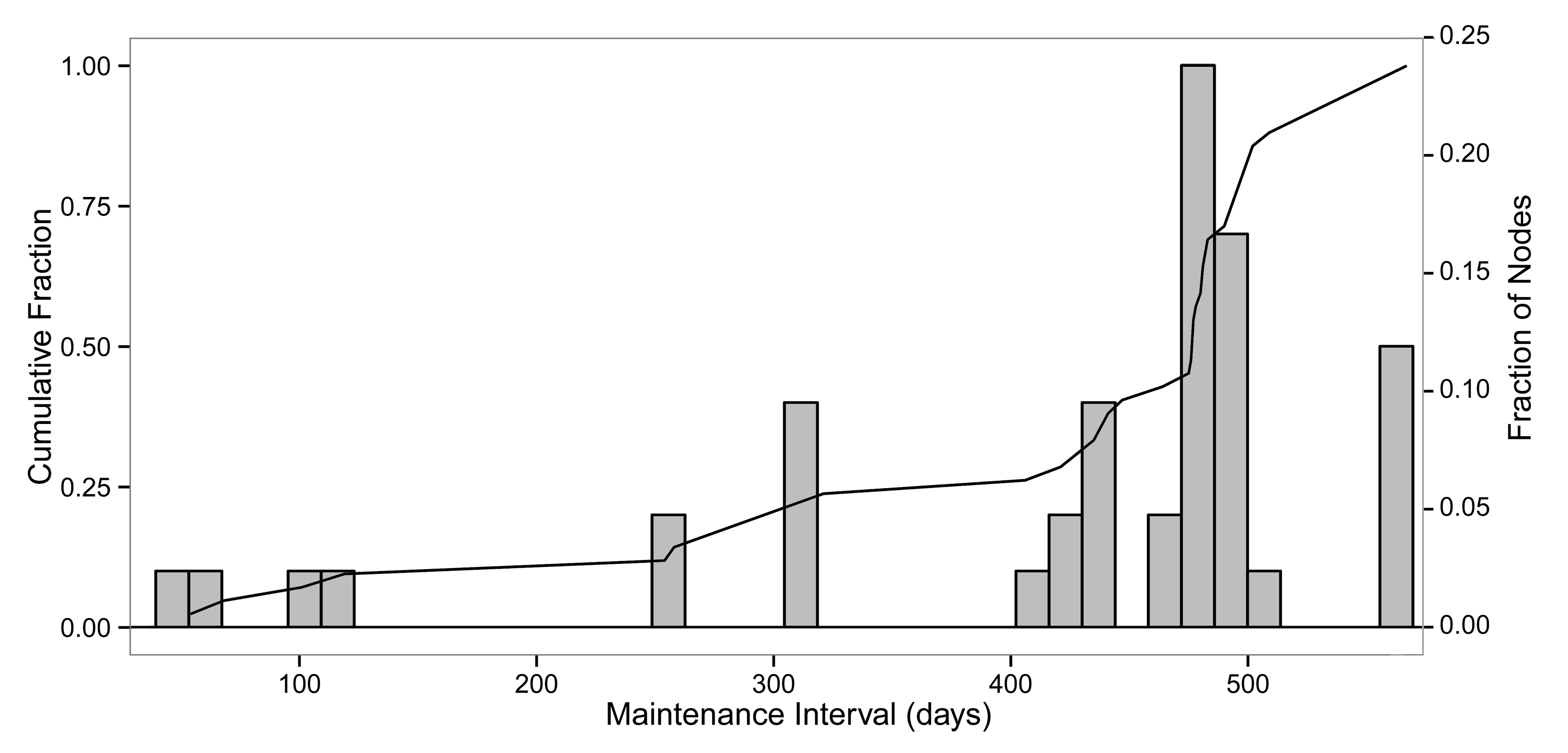}
\caption{\small Mote maintenance statistics of the Cub Hill deployment following site-wide battery replacement on 06/24/10. Solid line is the cumulative distribution of the node of maintenance intervals. The x axis indicates the number of days from installation of a node to the last date where 90\% or more of its expected data volume was recovered.  }
\label{fig:maintenance}
\end{figure}

Figure \ref{fig:maintenance} shows the distribution of maintenance intervals on our system. While a handful of devices fail within the first few months (likely due to physical/moisture damage), 90\% of the nodes ran for more than 6 months without maintenance, and 77\% ran for over a year without maintenance. With regular twice-yearly site visits, such a deployment could be expected to continuously deliver 90\%+ yields.

We evaluated mote failure according to the underlying causes. Software defects caused 11\% of the mote failures. About half of the failures were due to low batteries or moisture in their enclosures (43\% and 9\%, respectively). The rest were marked as ``unknown'', because they showed neither low battery nor high moisture before their disappearance. These could be nodes with a catastrophic leak during a rain event,  nodes that froze and drained their battery in one shot, nodes that suffered physical damage (such as a falling tree or animal chewing), or nodes that failed for no apparent reason.  Given that the deployment was in an urban environment, surprisingly, no motes were stolen or vandalized.


\vspace{4pt}\noindent
{\bf Wireless Frequency Band}
\vspace{4pt}

\noindent
First and foremost, as we have learned the hard way, the standard 2.4-2.5 GHz band is substantially absorbed by water vapor, and on days with a lot rain and humid air the range of the 2.5GHz radios have dropped substantially. So with our Gen 3 hardware we have switched to the 900MHz frequency band, and these problems have gone away.

\vspace{4pt}\noindent
{\bf Waterproofing}
\vspace{4pt}

\noindent
Even though the assembled boxes initially provided a perfect seal, the changing environmental conditions naturally led to increasing moisture. With the on-board moisture sensor, we were able to monitor the conditions inside the boxes and have seen the moisture steadily increasing. Temperature inside the box became quite high, sometime reaching 70$^o$C, due to sun exposure (greenhouse effect). This led to a substantial over-pressure inside, which over time created small micro-cracks in the gasket and the sealant, resulting in an equalization of the pressure. Once the weather changed and suddenly cold rain fell on the box, the temperature and pressure of the box dropped, sucking in moist air. Thus, during every cycle the moisture kept steadily increasing.  After about 6 months to a year this has caused the motes to stop working.  In the end, we have decided to use semi-sealed acrylic tubes with an open bottom, which are always in a temperature and humidity equilibrium with the outside.

\vspace{4pt}\noindent
{\bf Power Usage}
\vspace{4pt}

\noindent
Battery consumption was in important factor, and the typical analog sensors required a current draw of about 10mA.  In environmental science the can afford a modest data collection rate, so in practice we ended up using a 10ms/sensor sampling interval, taken at every 20 minutes. Even with 8 sensors this only amounts to about $100\mu$A for an 9-sensor mote, and about $25\mu$A for a mote with 2 sensors. 

Another lesson we learned that operating the motes at 3.3V leaves too small margin for the battery depletion. Due to temperature variations, the voltages of the LiPo batteries fluctuate in excess of 100 mV during a typical daily cycle. The dropout voltage on a typical voltage regulator does not leave much headroom for operating at 3.3V, so we have gone to 3V operating voltage and a very low dropout voltage chip, giving  us a much longer battery life.

\vspace{4pt}\noindent
{\bf Deployment Tradeoffs}
\vspace{4pt}

\noindent
We have also learned that requiring a full peer-to-peer network with real-time Internet connectivity so that we can assess the sensors' status at any moment makes practical deployments much too complex. While this was used for our big campaigns, e.g. Cub Hill in Baltimore, many other use cases require just a few sensors but spread out in many small clusters over a wide area. The price of sensors come into play as well. In the dense deployment we had many CO$_2$ sensors, with a typical cost of a few hundred to a few thousand dollars -- it was good to know that they were in place. The dense deployment use case was big and expensive enough that we have permanent internet gateway, with access to line power and a cable connection. 

The wide and shallow deployments typically require only a few temperature and moisture sensors, placed at various depths, typically 5-10-15 cm.  The moisture sensors are typically around \$100, and thus deploying them on a dedicated mote with a miniToast temperature sensor is a reasonable deployment choice.  In this setup we can operate the motes in a data logger mode, waking up the whole network about only once a month and using the mesh communication to incrementally download the new data. This mode saves power, thus enabled us to use even smaller batteries, making the whole assembly much smaller.

\vspace{4pt}\noindent
{\bf Software infrastructure}
\vspace{4pt}

\noindent
It was also extremely important that we adopted the well-tested TinyOS platform even when we moved to build our customized Gen 3 hardware. This was the reason why we stayed with the MSP430 based chip, but with the embedded multi-band radio, enabling us to go to 900MHz.

\section{Soil Moisture Results }

At the end of the day, what matters is whether these systems can provide insight into physical systems that would otherwise be impractical to attain. The rich dataset allows for a variety of analyses. For instance, we can compare conditions in different land use-land cover types using daily or monthly means and seasonal trends (Figure \ref{fig:irene}). Interestingly, monthly mean values in forest and grass were very similar, while values in the “other” category were consistently lower. 

\begin{figure}[ht]
\centering
\includegraphics[width=0.85\linewidth]{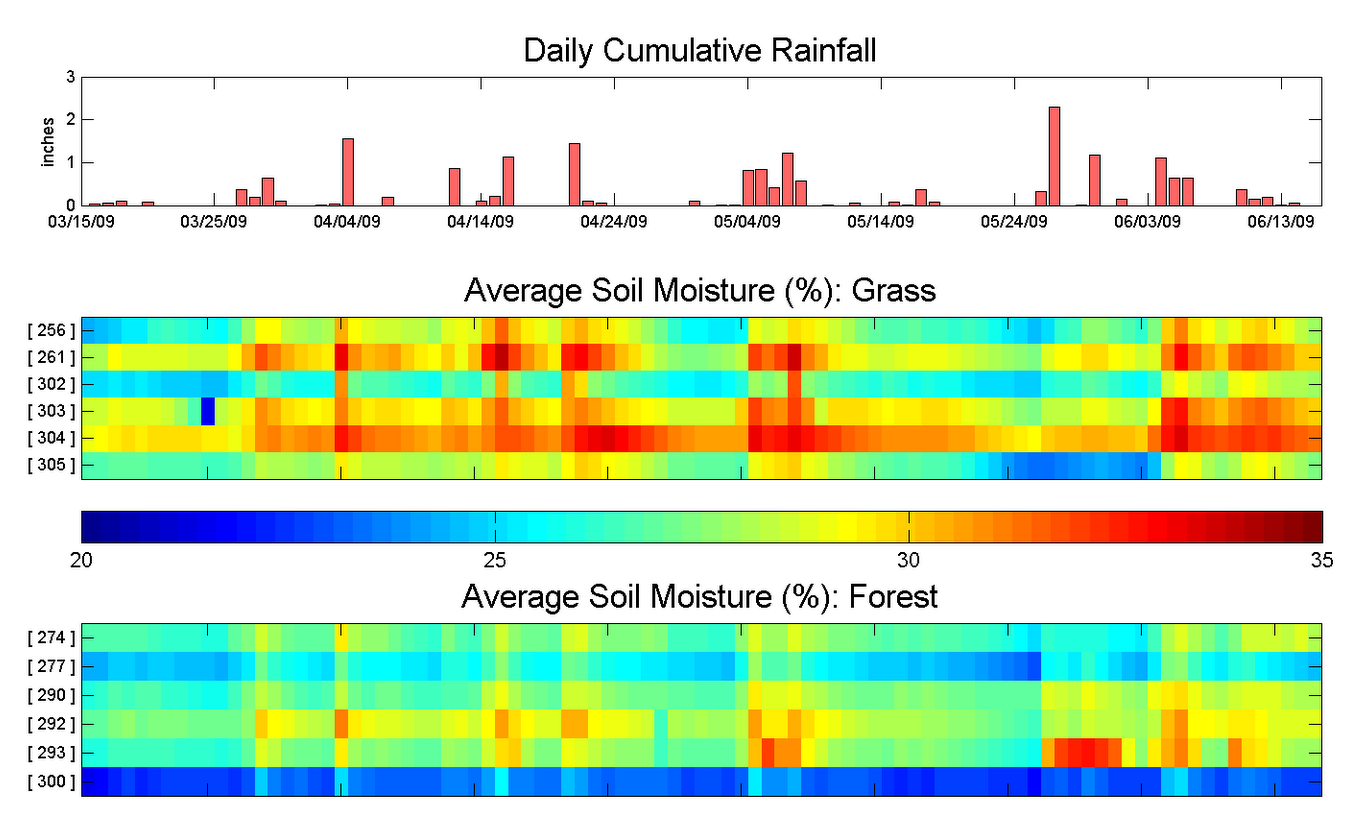}
\caption{\small Local variation of soil moisture (VMC) at two land cover types in the Cub Hill residential neighborhood in Baltimore County, MD, between March and June 2009.  A subset of locations are shown. Each cell represents daily average values at 10 cm depth. Daily precipitation values are also shown. }
\label{fig:moisturecolor}
\end{figure}

\begin{figure}
\centering
\includegraphics[width=0.9\linewidth]{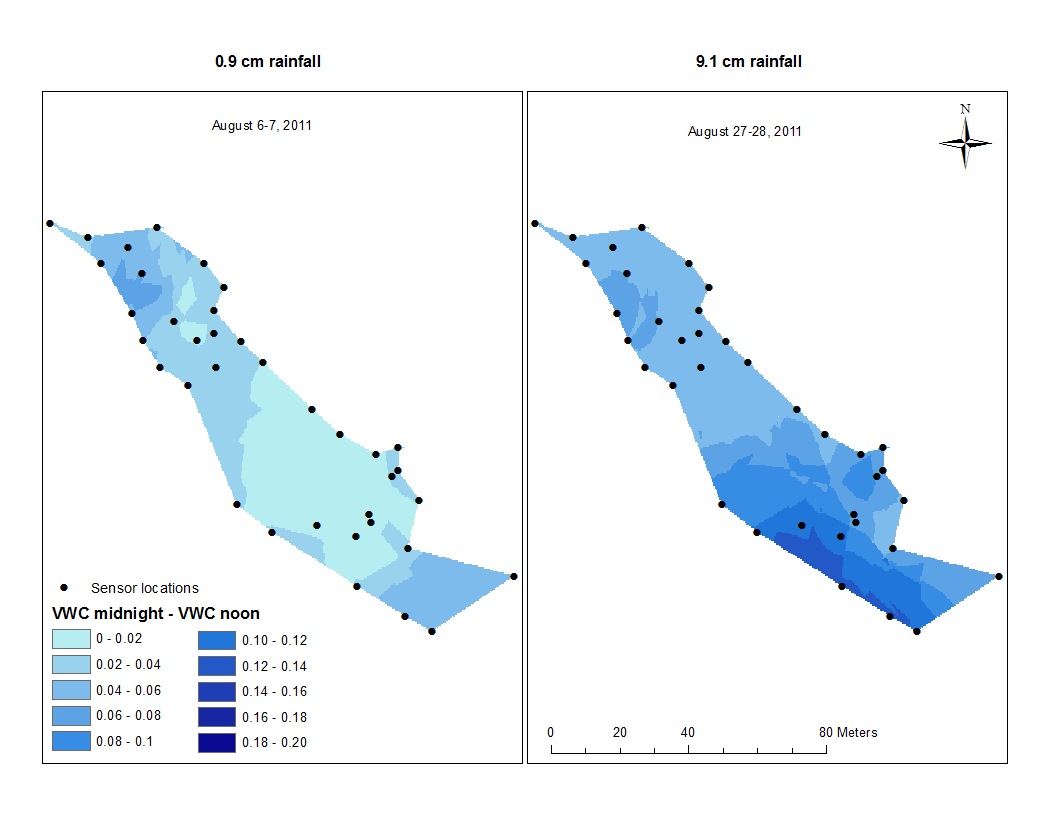}
\caption{\small Response of soil moisture to extreme precipitation at Cub Hill. Spatial map shows the change in volumetric soil moisture content between hours before rain and during peak precipitation intensity.measurements were taken at 10 cm depths.  A: regular rain event, B: Hurricane Irene. }
\label{fig:irene}
\end{figure}

Figures \ref{fig:moisturecolor} and \ref{fig:irene} illustrate what makes WSNs so well-suited for this domain. Even though the averages are similar, the individual sampling locations varied a great deal. Moreover, our dataset shows how soil moisture responds to rainfall events at fine temporal resolution, grouping locations by ground cover. Not only can we see clear differences between the cover types, we can capture the small-scale heterogeneity between sampling locations under the same cover. More detailed analysis of these data and the interactions on soil temperature and soil moisture are discussed in \cite{Savva12}. Spatial analysis allows us to observe the soil response to extreme events (Figure \ref{fig:irene}), to explore temporal stability of spatial structure \cite{Savva12}, and to interpret soil fauna distribution \cite{Szlavecz11} in this residential neighborhood. Soil moisture data, along with continuous soil CO$_{2}$ concentration data collected by the WSN can be used to build a model for soil CO$_{2}$ efflux in urban forest and grass \cite{Chun11}. Combined with the urban CO$_{2}$ flux tower data this information is then used to build a coupled urban carbon and water cycle model (John Hom, pers. comm.).

The primary objective of the Life Under Your Feet project was to design and build an end-to-end wireless sensor network (WSN) for monitoring the soil ecosystem and to test the system in a variety of environmental conditions. Similar efforts have been reported \cite{Cardell05,Sikka06,Ramanathan09}, or still ongoing (http://soilscape.eecs.umich.edu/). To our knowledge, at the time of the deployment, the Cub Hill project was the largest and it was the longest experiment of its size. In general, our deployments met their key goals: they survived for long periods in the field, while reliably collecting large volumes of data. The data can be utilized for a variety of projects from building hydrology models to identifying biological activity hotspots. A wireless sensor network can be used for validation of remotely sensed soil moisture data. As technology advances, the system keeps evolving. However, further development of hardware will always have to optimize among conflicting demands: increasing scale of deployment, while keeping the system reliable, and keeping the cost low. The Gen-3 platform and software addresses many of the lessons we learned from previous deployments, and the cost of one mote (\$20) is now negligible compared to the cost of the sensors themselves. Ultimately the goal of such approach is to gain insight to the spatio-temporally complex and extremely species rich soil ecosystem. 

\section{The Gen-3 Hardware}

\vspace{4pt}\noindent
{\bf The Bacon}
\vspace{4pt}

\begin{figure}[ht]
\centering
\includegraphics[width=0.65\linewidth]{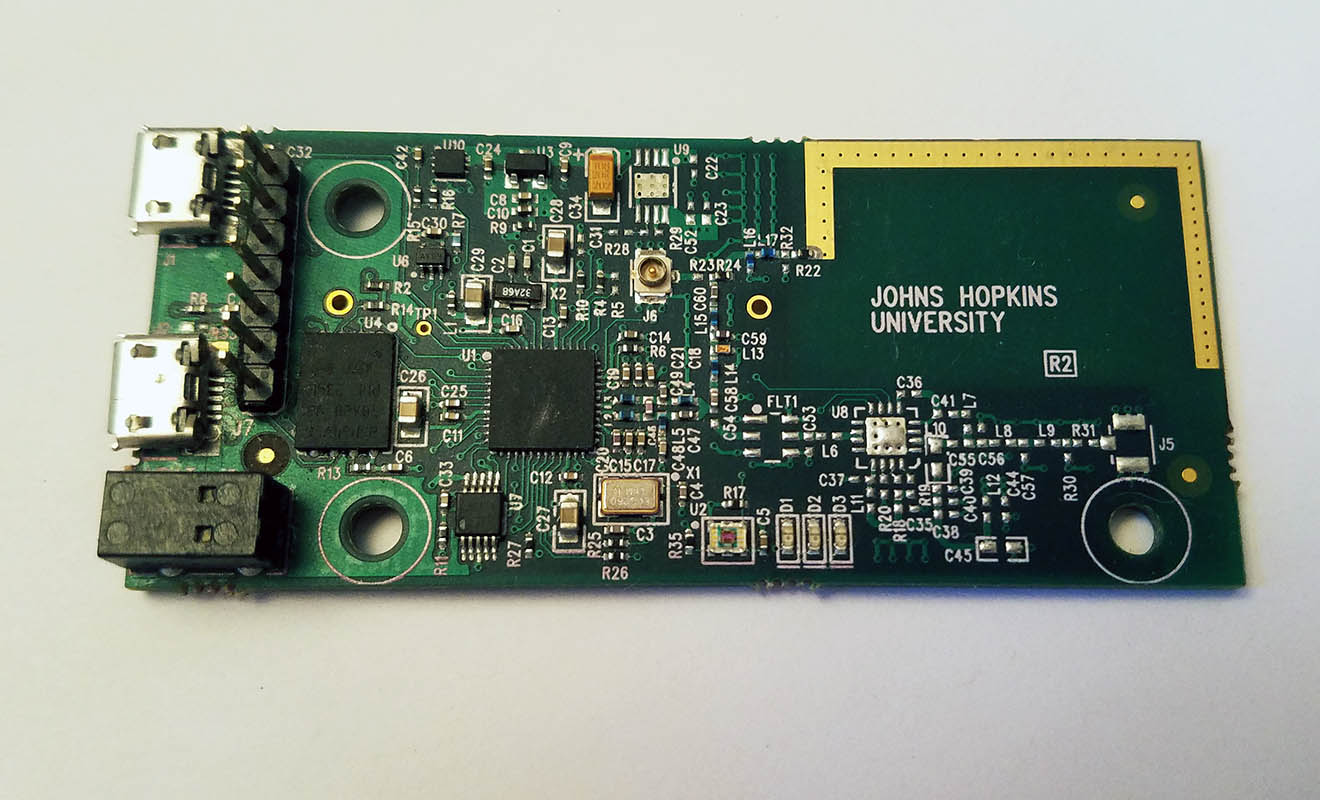}

\caption{\small The CC430-based Bacon node, our Gen-3 radio platform.}
\label{fig:bacon}
\end{figure}

\noindent
The Breakfast Suite has several modular components. The main unit is the so called Bacon board, which has the wireless transmitter.  We decided to use the CC430F5137IRGZ MCU, which has several useful properties. It has the MSP430 core, immediately providing compatibility with the existing TinyOS stack. It has a multiband radio transmitter supporting the 900Mhz band, it still uses the ZigBee packet format on all frequencies, has more processing power and eight times more memory than the original MSP430 (Figure \ref{fig:bacon}). Furthermore, it has ultra-low power consumption. We have also placed a 64 Mbit serial flash memory on board (Numonyx M25P64-VME6G) for buffering the acquired samples.  We are using a low dropout LDO providing 3V of stabilized voltage (MCP1700T-3002E/TT, with a dropout voltage $<$ 25mV at a 25mA current). 

The board has options for an on-board 1/2 AA 3.7V LiPo battery as well as it has an external battery connector. The boards have been also designed to accept a radio power amplifier, and an external antenna, but we have only populated about 100 boards with that option. These motes can serve as longer range relay nodes. 
We have an on-board analog switch turning the flash power off when it is not required, saving additional current in deep sleep mode. We have an additional 10-pin JTAG connector, for deep debugging of the system.

We have two sensors on board, one is a linear thermistor (Microchip	MCP9700AT-E/LT) running in 12-bit, 0.1 $\deg$C sensitivity, the other is an Avago APDS-9007 ambient light sensor. We are using two micro USB connectors, one to connect to a programmer board, the other to connect to the multiplexers.  The board is also actively monitoring (and saving) the battery voltage.

\vspace{4pt}\noindent
{\bf The Toast}
\vspace{4pt}

\noindent
The Toast board is an intelligent multiplexer/interface for up to eight analog sensors. It is based on a light-weight MSP430F235TRGCR MCU, which takes commands from the Bacon node, turns the sensors on for measurement for a predetermined period (10ms), acquires an analog voltage measurement for each enabled channel and uploads the data to the Bacon for storage in flash.  The communication between the Bacon and the Toast is using the I2C protocol. Several Toast modules can be daisy chained.

\begin{figure}[ht]
\centering
\includegraphics[width=0.65\linewidth]{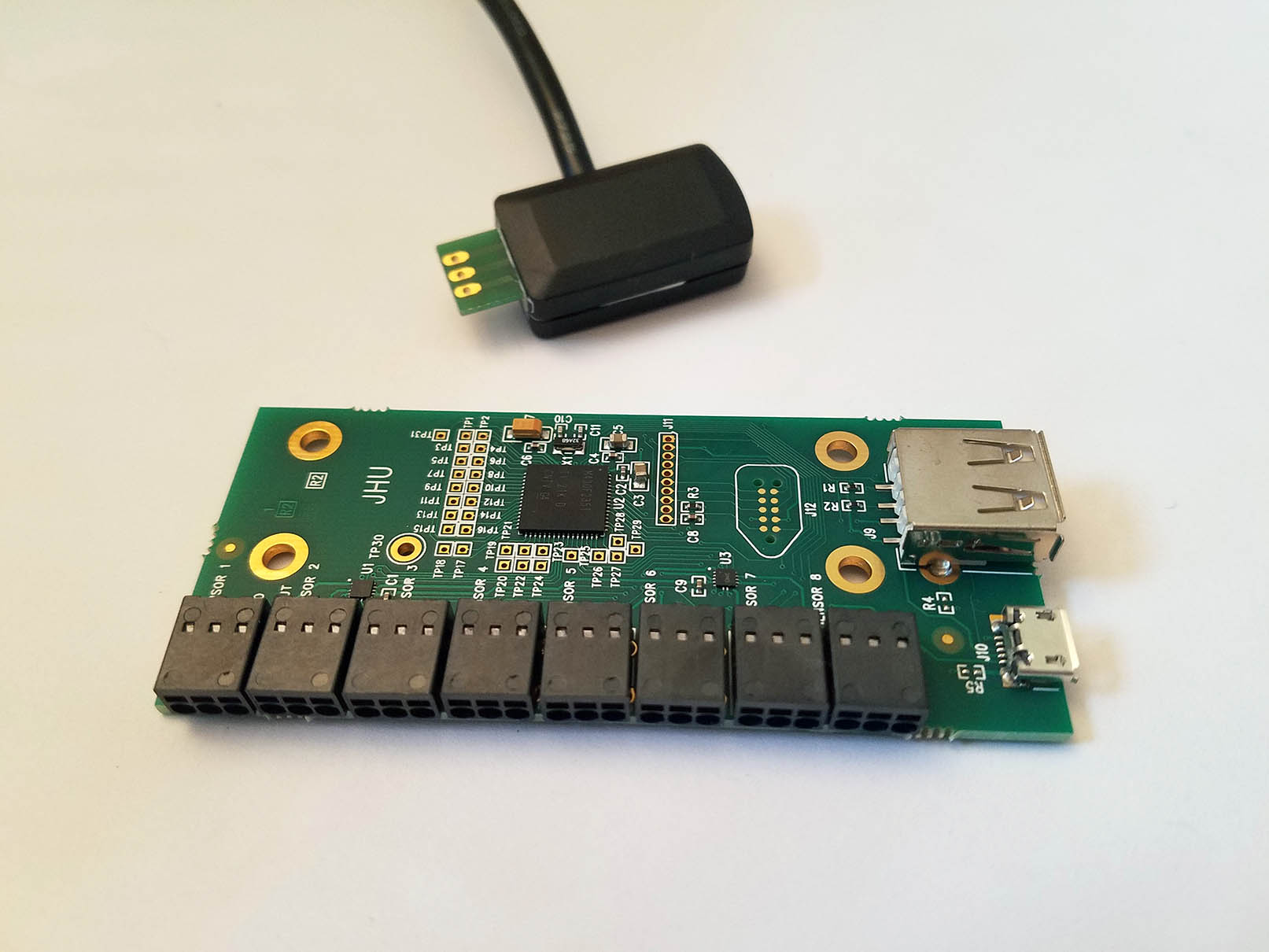}
\caption{\small The CC430-based Toast and miniToast analog interface/multiplexer boards.}
\label{fig:toast}
\end{figure}

\vspace{4pt}\noindent
{\bf The miniToast}
\vspace{4pt}

\noindent
The miniToast is a compact, simplified version of the Toast for simple mote configurations, when only a single analog sensor is needed. It includes two additional internal channels, one is a high precision analog thermistor (US Sensor PS103J2), the other is a measurement of the supply voltage to the sensors.
The miniToast module is placed in epoxy so that it can be buried underground, together with the additional external sensor. 

\vspace{4pt}\noindent
{\bf The USB Programmer}
\vspace{4pt}

\noindent
This board enables the programming of the Bacon and Toast boards from a PC host.
It contains an FTDI232R interface chip, and an ADG715T 3.3V voltage regulator.
For programming the Bacon and Toast combination, we need to first connect a Bacon board to the programmer, and then the Toast. One can update the firmware on each boards, as well as add id numbers (barcodes) to both devices as well as for the individual sensors. These ID numbers are than contained in the various status messages and data records.

\begin{figure}[ht]
\centering
\includegraphics[width=0.65\linewidth]{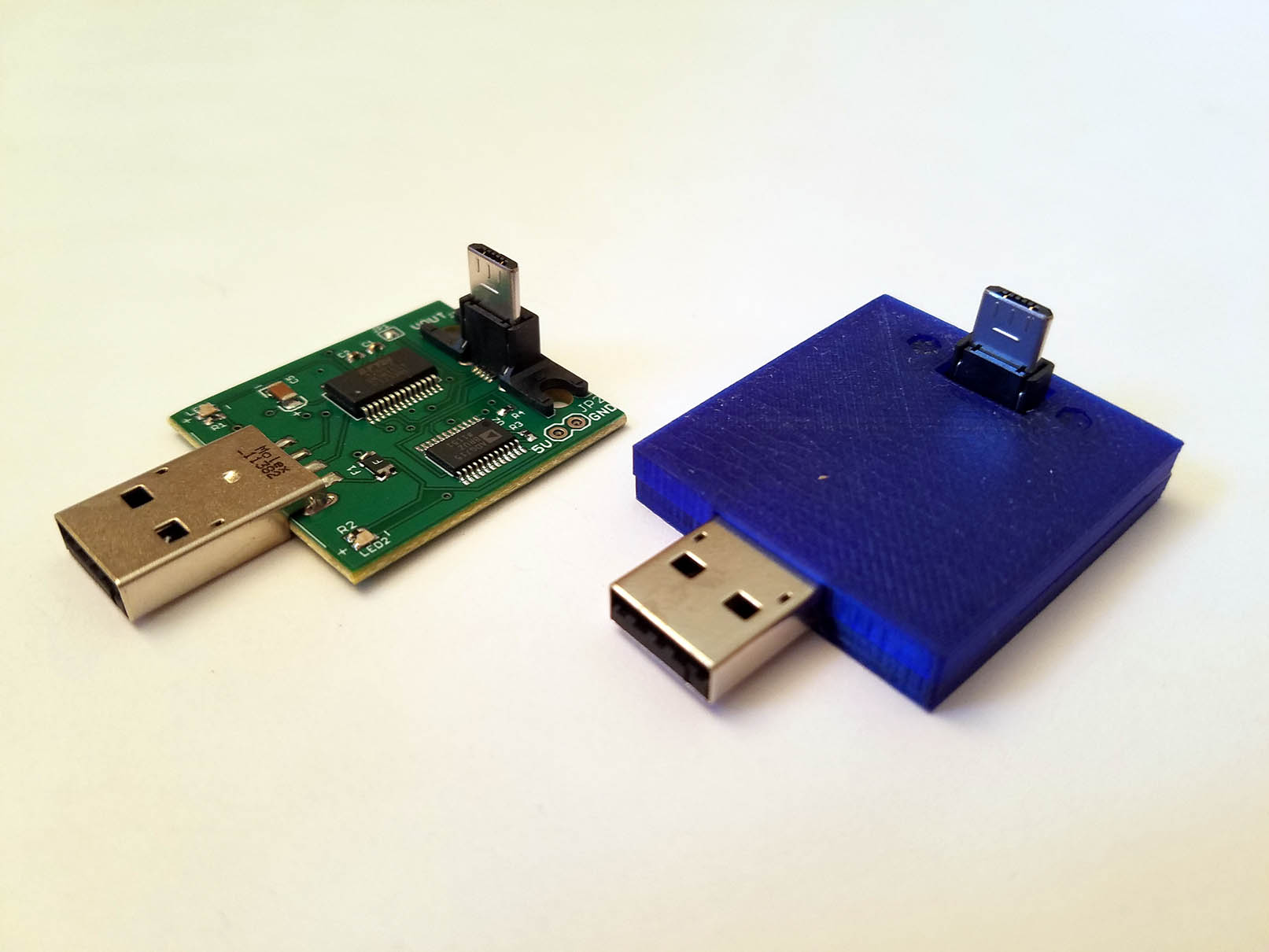}
\caption{\small The USB programmer board, with and without its 3D-printed enclosure.}
\label{fig:usb}
\end{figure}

\section {The Gen-3 Software Stack}
{\bf Concurrent Transmission w. Forwarder Selection}
\vspace{4pt}

\noindent
One major goal of the Gen-3 platform is to to avoid the inherent limitations in selecting data transmission routes over unreliable hardware and fickle link dynamics. The Koala data collection system\cite{Musaloiu08} uses a central source-routing mechanism where the base station decides the route of each download by analyzing the probes sent from nodes in the network. However, when the batteries of a few nodes in the network are depleted, the base station often fails to find a reliable route, since a node on a low voltage may still send wake-up probes and be included in the forwarder route, even if it could no longer reliably transmit data. Consequently, frequent download retries cause high duty cycles that drain the batteries even faster. One possible solution to the problem is to use detailed node status information (e.g. battery voltage, humidity in the enclosure) in the route selection process, yet obtaining detailed link quality information demand significant energy and computation costs, placing a high burden on resource-constrained motes.

In Gen-3 network, we take an alternative method by leveraging non-destructive interference to create redundant simultaneous delivery paths to multiply the probability of reaching the end node over an unreliable network. Under this approach, we use simple hop counts to identify a subset of network that roughly lies between the source and destination as the set of potential forwarders. By using the precisely-timed radio on board bacon motes to schedule transmissions to occur simultaneously, we can send a data packet over this set of nodes with very low risk of loss due to interference. This protocol, which we call concurrent transmission with forwarder selection (CXFS), requires neither the overhead of calculating a single optimal path between the the source and the sink nor the costs of indiscriminately forwarding the message across the entire network as in a simple flooding method. This balance between simplicity and selectivity helps the sensor network attain both high throughout and low energy consumption. 

\vspace{4pt}\noindent
{\bf Multi-tiered Network Hierarchy}
\vspace{4pt}

\noindent
Studies in soil ecology sometimes require comparing distinct habitats over a relatively large region while simultaneously capturing the heterogeneity within each habitat. This configuration often leads to the need of setting up multiple sensors in small clusters. To account for the spatial patchiness of the deployment, the entire deployment is divided into one or more patches, where each patch contains a router mote equipped with long range transceivers and a cluster of leaves motes. A router independently collects data from the leaves in its patch, and a basestation periodically downloads data from all routers. This multi-tiered structure avoids the need to install multiple relay motes between patches and allows energy-efficient sampling over heterogeneous landscape.

\begin{figure}[ht]
\centering
\includegraphics[width=0.7\linewidth]{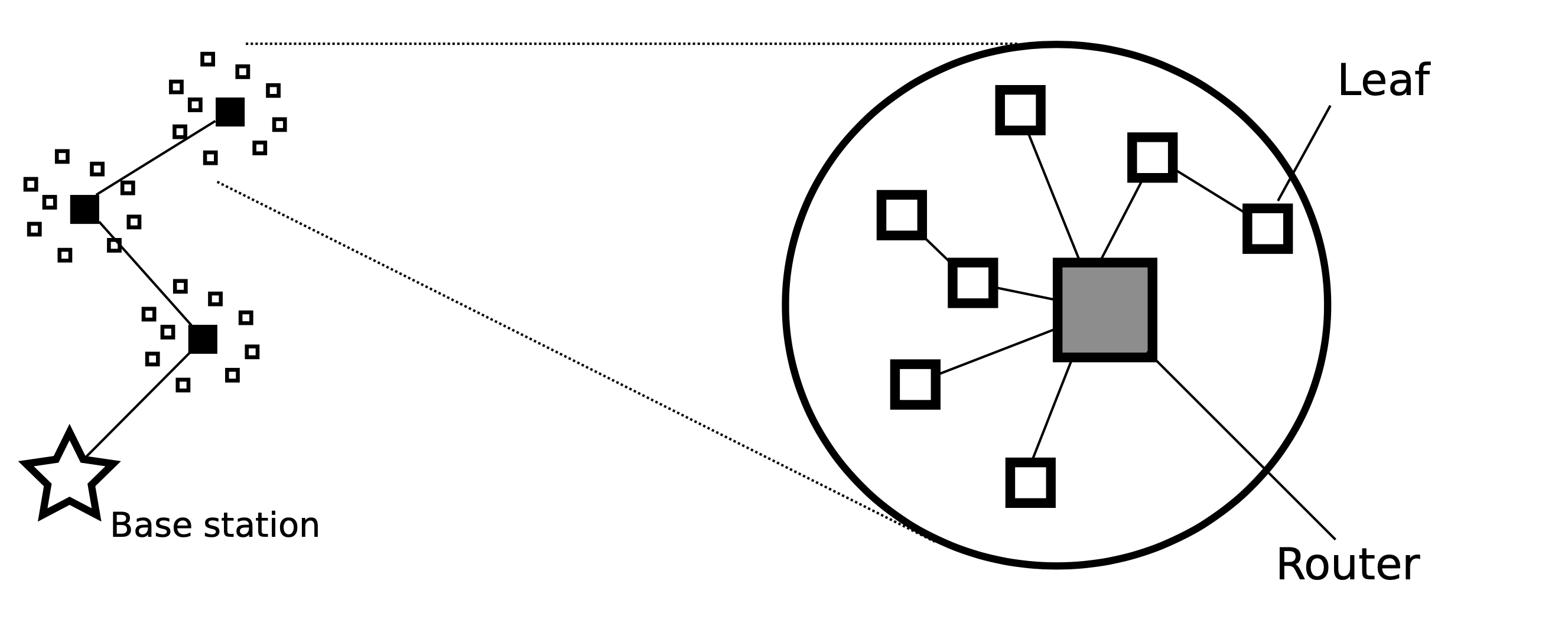}
\caption{\small Multi-tiered network segments.}
\label{fig:hierarchy}
\end{figure}

\vspace{4pt}\noindent
{\bf Data Pipeline and Meta Data Management}
\vspace{4pt}

\noindent
With a hardware interface to analog sensors and a reliable networking structure in place, the next step is to build an end-to-end data pipeline from sensors to database. On each mote, we divide the flash storage into a setting storage and a log records storage. The former is used to store customized configurations such as sampling interval and unique id, whereas the later is used by both the Bacon and Toast modules to record samples collected. An auto-push component keeps track of data recorded in the flash, pushes the data from the flash to the network stack when needed, and handles recover requests for missing data based on specific cookie and length parameters. If cases where log record packets are forwarded from leaf motes to a router mote, they will be temporarily stored in the router's log with a TUNNELED\_MESSAGE record appended. During a download, the base station will first attempt to receive all information from the router's log, and it will retrieve log records from the leaf motes only if the information is absent from the router.
\begin{figure}[ht]
\centering
\includegraphics[width=0.85\linewidth]{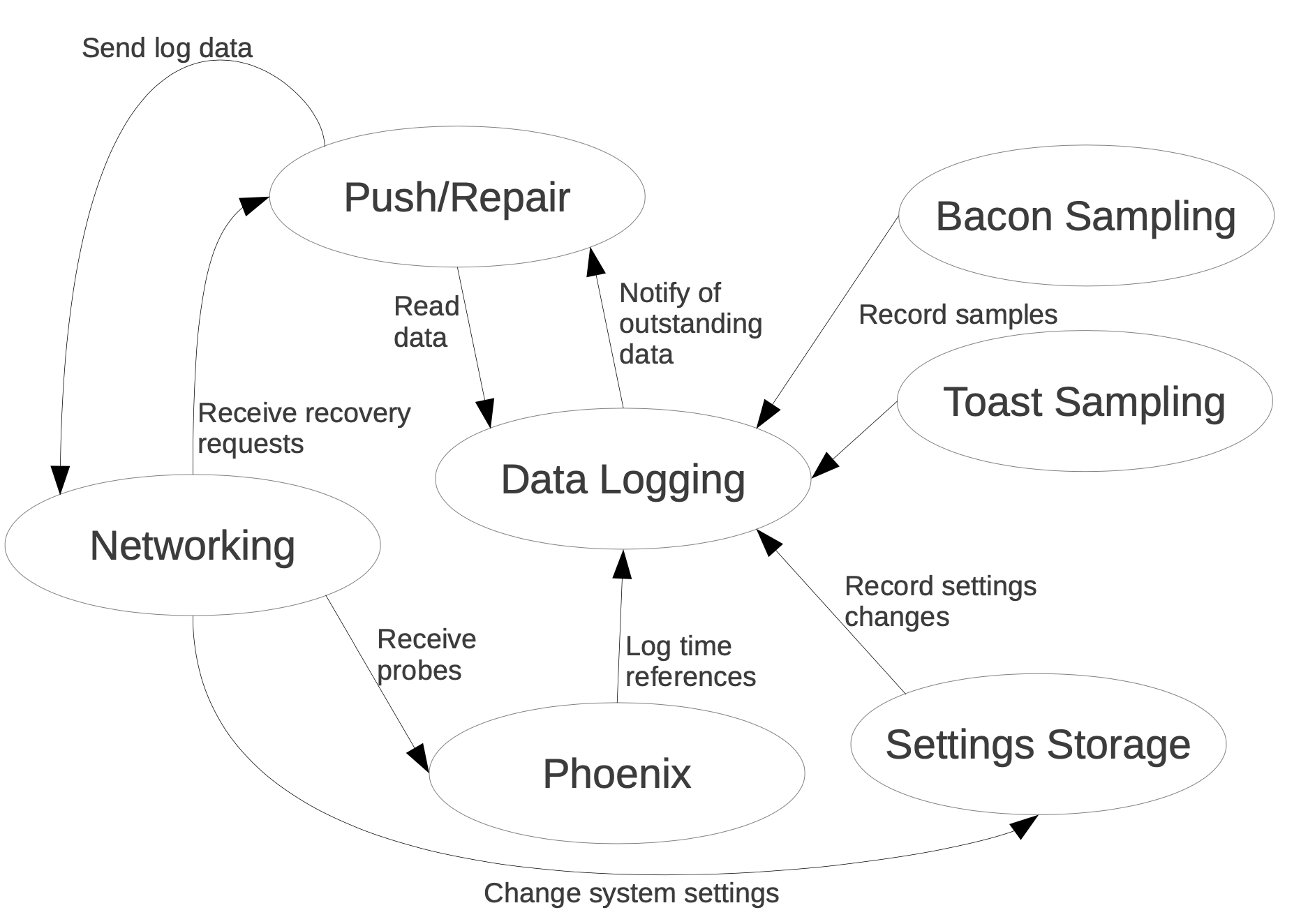}
\caption{\small Interactions between major components of the software.}
\label{fig:components}
\end{figure}

A Python program on the PC will process each log record packet when they arrive from the base station. The program first stores a safe copy of the raw packet in a binary table, and then parses the packet to extract individual records. Each record is identified by its type and dispatched to a corresponding decoder. The decoder then decodes the record to an ASCII string, extracts the fields from the string, and inserts the fields into a matching table in the database. 

The above description highlights the flow of data from the sensor to the database. Yet for the system to collect scientifically meaningful data, it is also critical to systematically track deployment metadata such as the type of the sensor, the channel assignment, and the pairing between bacon and toast boards. In the third generation software, metadata tracking is enforced at installation stage by a dedicated labeler program. The manufacturer ID, barcode ID, and sensor assignments are stored in both a local database and the setting storage volumes of the Bacon and Toast boards. The program will not generate data reports unless the metadata associated with the deployment is present.

\begin{figure}[ht]
\centering
\includegraphics[width=0.9\linewidth]{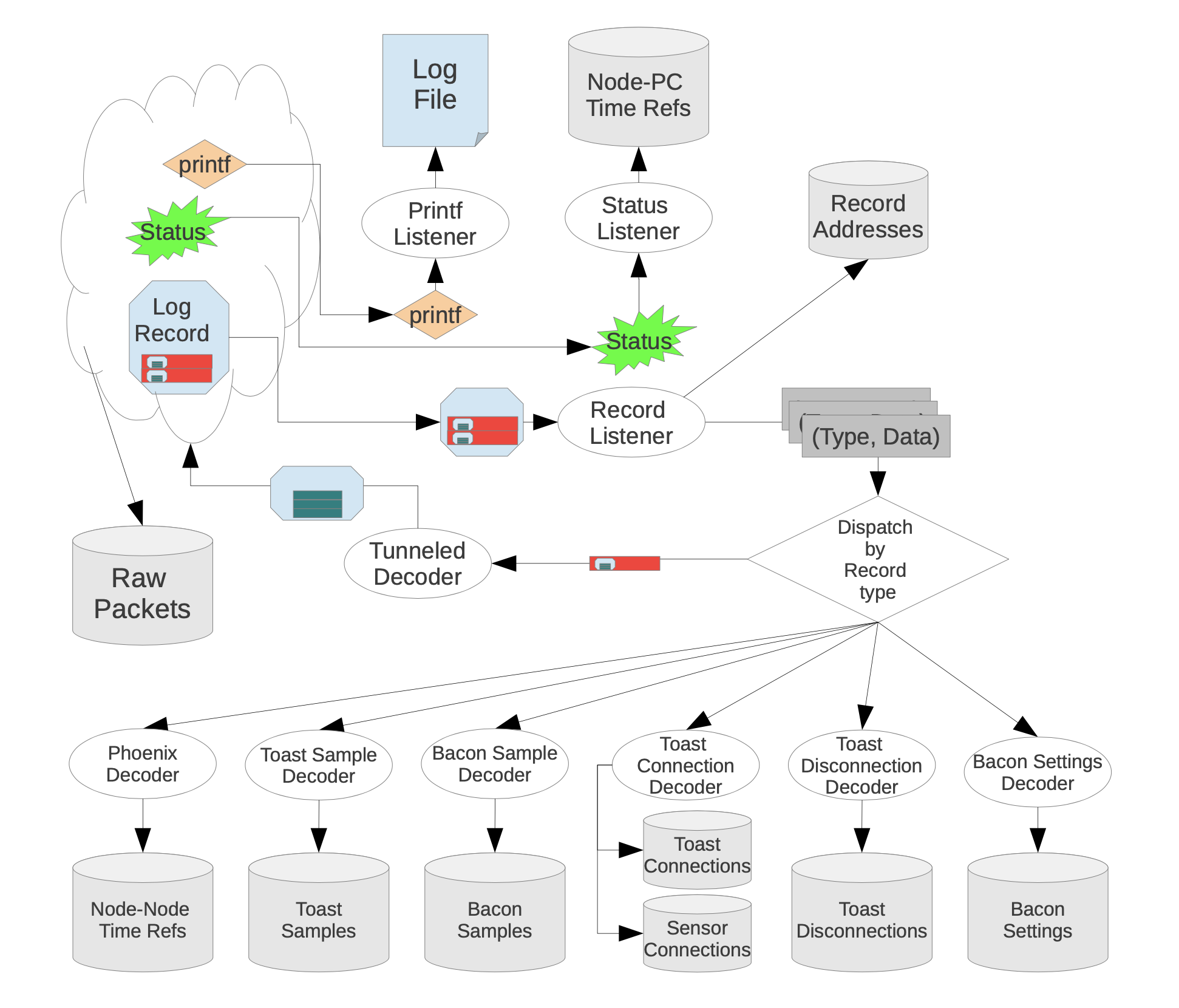}
\caption{\small Data flow in PC.}
\label{fig:dataflow}
\end{figure}

\section {The Gen-4 Software Platform}
{\bf Integrated GUI }
\vspace{4pt}

\noindent
The completion of the Gen-3 platform enabled a wide range of environmental monitoring opportunities. After the development, the system has been test-deployed in SERC and yielded promising scientific data. Despite the technical success, our field tests revealed several stability issues that required a consistent involvement of technicians, and the platform’s operational complexity hindered its wide adoption by soil scientists. The current interface only allows a user to download all the data from a patch of sensors, yet a user may be interested in retrieving data from a specific mote to obtain data with higher granularity. Furthermore, the Gen-3 system does not perform quality-control measures on the data collected or alert the user of any potential issues, so it might take an end-user several months to realize the issue while the critical window of research is wasted. 

\begin{figure}[ht]
\centering
\includegraphics[width=0.80\linewidth]{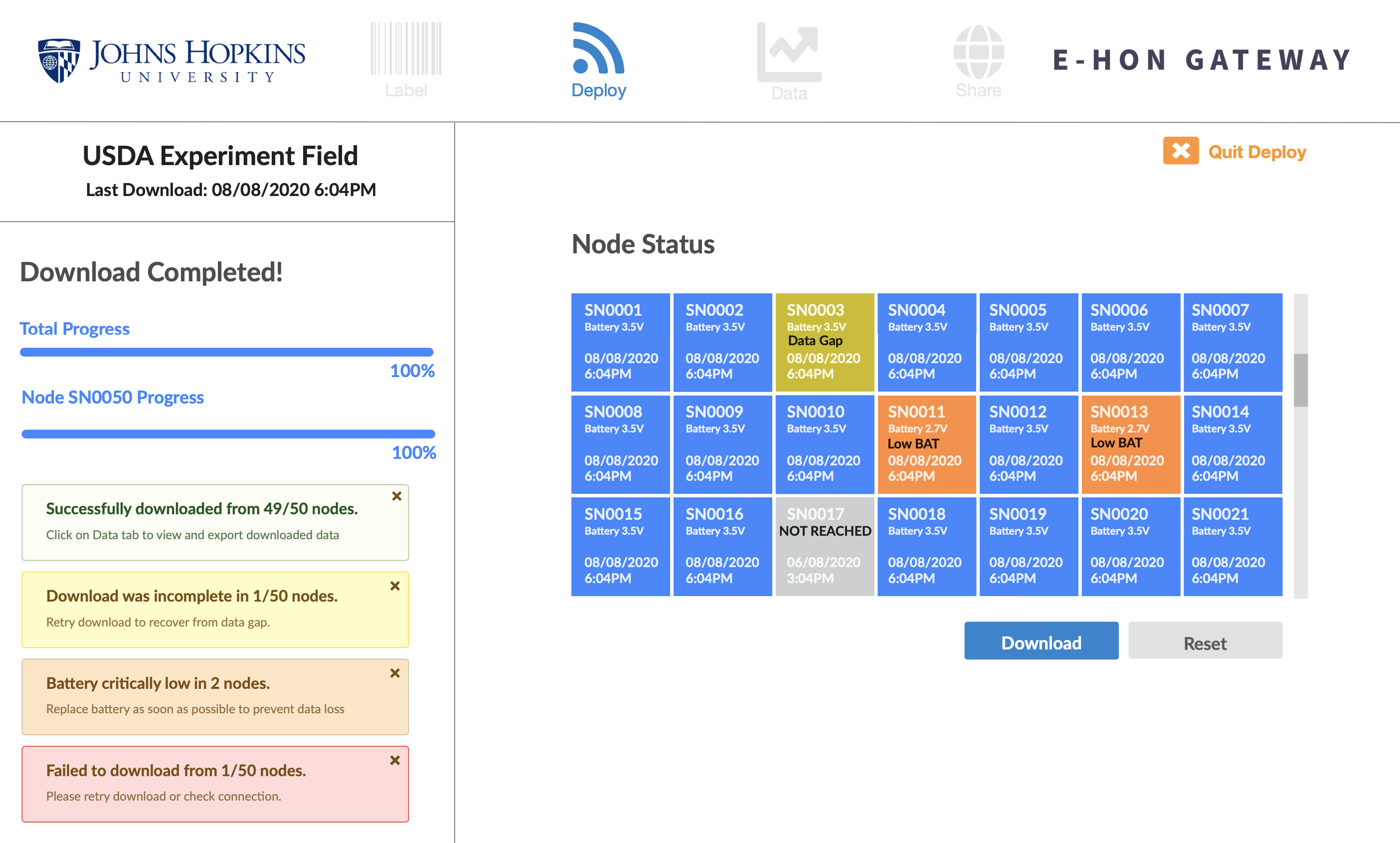}
\caption{\small Prototype design of the generation-4 software platform. The new software will provide a convenient interface for users to interact with the sensor network}
\label{fig:interface}
\end{figure}

These limitations have prompted us to redesign the high-level, user facing components, and to create an integrated software platform to streamline device management, field deployment, quality control and data reporting. A recent Python-based web-app framework (Dash) provides us the ability to implement such a platform with comparative ease. Through a web-based graphic interface, the user will be first guided through the process of labelling motes and sensors, setting radio channels and sampling intervals, and managing deployment-wide metadata (Figure \ref{fig:interface}). When in field, the user can manage the deployment, edit the metadata for each unit, or add or remove a device if needed. Through an interactive portal, the user can download data from the entire deployment, a patch within the deployment, or a specific mote as desired. Progress bars and a color grid on the screen will display the status of download. When a download completes, the status of each mote,  including battery voltage and percentage of samples retrieved, will be reported. The platform will then run a quality control algorithm to check the database for missing, duplicate, or anomalous data, and alert the user if the problem is not recoverable by automatic correction. This platform will also provide customizable data visualization and report functions, allowing a user to generate spreadsheets and graphics with only a few clicks ``in situ''. Finally, the platform will also provide optional user authentication and data upload function, allowing a user to share data while controlling access.

\vspace{4pt}\noindent
{\bf Data Hosting and Analytics}
\vspace{4pt}

\noindent
Our data will be hosted on a scalable collaborative data analytics platform, which integrates about 6PB of file storage, several PB of databases, automatically generates user logs, and provides about 100 virtual machines for interactive and batch computations. All code, database tables and data products are shareable at different granularities (users, groups, world). The system today is supporting more than 70 different projects in a variety of science domains, from astronomy to turbulence. The analyses can also use server-side Jupyter/iPython notebooks. These are preconfigured with database access, thus users can run their SQL queries out of Python. The Jupyter environment also enables Matlab and R. All major machine learning toolkits are available on the system, including several V100-based GPU servers. Most of the sensor data collected so far has already been moved to the SciServer, to alleviate the need for maintaining a separate data system for our project. We will use this system for all future data hosting and analysis.

\section{Summary}

Soil moisture is a key driver of many soil physical and biogeochemical processes, and an important component of the water cycle. In the US several platforms exist to monitor soil moisture, but a collective effort to establish a National Soil Monitoring Network has emerged only recently. Currently neither the existing networks nor the SMAP satellite provide data on soil moisture in urban environments. Due to the enormous heterogeneity of the urban landscape, the diversity of urban soils from remnants of naturally developing soils to engineered substrates, and various management practices, physico-chemical properties cannot be simply derived or modeled from natural soil forming factors and thus have to be measured directly.

An ideal urban soil monitoring network would consist of many small sensor clusters, and should deployed on all major land use types. Lawns parks are a permanent feature in the urban-suburban landscape, but additional land uses, vacant lots, community gardens, remnant patches of regional biome, and other unmanaged areas should also be monitored. Wireless data collection allows less disturbance and less intrusion of the sites, which are often privately owned. 

In Baltimore, in our next major deployment project, we are planning to monitor over 200 sites, called ``parcels''. A parcel is a unit of investigation and it is managed by a single entity: a company, organization, or an individual. The plan is to deploy a set of soil temperature and moisture sensors at different depths, and in several patches within the parcel, for example lawns and planting beds. 

There is an increasing need for this type of information not only in the scientific community but in various sectors of the community. For instance, cities are relying on, and even by policy requiring, alternative stormwater management technologies, from green roofs, to living walls, bioswales, and bioretention systems. Yet, there is little opportunity for design and engineering professionals to evaluate their post-construction performance in order to improve future projects and revise sustainable design guidelines. Sensor technology would enable designers, engineers, and scientists to collect the data required to meet the needs of urban populations by improving the development high-performance landscapes and green infrastructure projects that are multi-functional in protecting environmental quality, while providing social and economic benefits in urban centers. 

Our Gen-3 platform can be ideally used for the wide area deployment described above. The main challenge for such wide area deployments is to make the whole configuration, deployment and ongoing data collection process much more streamlined and simple, and enable a variety of users, including soil scientists, urban ecologists, practitioners, as well as community scientists to be in full control of the whole experiment. In the same spirit, the analysis of the data must be made much simpler by using an integrated collaborative analysis environment, which integrates databases, file stores and Jupyter notebooks.

In this paper we presented the history of developing several generations of inexpensive sensors to measure soil properties. Our architecture evolved a lot over the years and now we have an inexpensive and robust hardware/software architecture that can collect data at an extremely low power consumption, and can be used in deployments such as the one described above. 

\section*{\uppercase{Acknowledgements}}

The Life Under Your Feet project started out with seed grants from Microsoft Research and the Seaver Institute. We are most grateful to Jim Gray who was instrumental in getting this project off the ground, and to Dan Fay and Tony Hey for their continuing support and encouragement. Later this project was partially supported by several grants from the National Science Foundation (NSF IDBR-0754782 and NSF DEB-0423476, NSF-ERC EEC-0540832). A seed grand from JHU 21 Century Cities Initiative supported the current research.  
The Gordon and Betty Moore Foundation sponsored the development of the Breakfast Suite. Undergraduates Josh Cogan, Julia Klofas and Justin Silverman helped building and testing the WSN. Mike Liang contributed to the software development; Jordan Raddick, Taesung Kim and Luis Grimaldo were instrumental in developing Grazor. Thanks are due to the Maryland Department of Natural Resources for allowing us to use their site as testbed, and to John Hom of the US Forest Service for hosting the gateway computer in his lab at Cub Hill.

\bibliographystyle{apalike}
{\small
\bibliography{reference}}

\end{document}